 \documentclass[final,5p,times,twocolumn,number,sort&compress]{elsarticle}

\usepackage{amssymb}

\usepackage{hyperref}
\hypersetup{colorlinks,urlcolor=blue,citecolor=blue,linkcolor=blue}

\journal{\href{https://doi.org/10.48550/arXiv.2310.06052}{arXiv:2310.06052}, published in \href{https://doi.org/10.1209/0295-5075/ad2ff7}{Europhys.Lett.  145 (2024) 69001.}}

\begin{document}

\begin{frontmatter}



\title{Constraint on Lorentz invariance violation from Vela pulsar}

\author[first]{Hao Li}
\author[first]{Jie Zhu}
\affiliation[first]{organization={School of Physics, Peking University},
            city={Beijing 100871},
            country={China}}
\author[first,second,third]{Bo-Qiang Ma \texorpdfstring{\corref{cor1}}{}}
\ead{mabq@pku.edu.cn}
\cortext[cor1]{Corresponding author}

\affiliation[second]{organization={Center for High Energy Physics, Peking University},
            city={Beijing 100871},
            country={China}}
\affiliation[third]{organization={Collaborative Innovation Center of Quantum Matter},
            city={Beijing},
            country={China}}

\begin{abstract}
The High Energy Stereoscopic System (H.E.S.S.) Collaboration reported 
the discovery of a novel radiation component from the Vela pulsar 
by their Cherenkov telescopes. 
It is of great importance that gamma rays with energies of at least 20~TeV are recorded unexpectedly. 
The H.E.S.S. Collaboration argued that such results may challenge the state-of-the-art models for the high-energy emission of pulsars.
We point out in this work that these results also provide 
a unique opportunity to constrain certain Lorentz invariance violation parameters, leading to the realization of 
studying Lorentz invariance violation by using gamma-ray pulsars.
The Lorentz invariance violation scale is constrained at the level of $E_{\mathrm{LV,}1}> 1.66\times 10^{17} \rm GeV$ for the linear scenario, and $E_{\mathrm{LV,}2}>3.53\times 10^{10} \rm GeV$ for the quadratic scenario.
We anticipate that digging into the detailed features of the data of the Vela pulsar and analyzing potentially more very-high-energy photon data from pulsars in the future would improve the constraints on Lorentz invariance violation.
\end{abstract}



\begin{keyword}
Lorentz invariance violation \sep the Vela pulsar \sep very-high-energy gamma rays



\end{keyword}

\end{frontmatter}




Utilizing high-energy photons from distant objects in the Universe is one of the most important approaches to searching Lorentz invariance violation~(LIV)~\cite{Amelino-Camelia1997a,Amelino-Camelia1998}.
In general, these sources are required to be at a very large distance from the Earth and the emissions of energetic photons are highly variable, with a time-structure of milliseconds~\cite{Amelino-Camelia1997a,Amelino-Camelia1998}.
Therefore, gamma-ray bursts~(GRBs) have long been the central focus for the study of LIV following this method~(see Refs.~\cite{wei2021tests,He:2022gyk} and references therein). There are also proposals 
that very-high-energy~(VHE) photons from pulsars can be used to study LIV~\cite{Otte:2012tw,Zitzer:2013gka,MAGIC:2017vah}.
Just recently, results from the High Energy Stereoscopic System~(H.E.S.S.) 
provide a unique opportunity
to searching Lorentz invariance violation from 
VHE cosmic photons.
The Cherenkov telescopes of H.E.S.S. detected multi-TeV gamma-rays from the Vela pulsar, PSR B0833045, of which the radiation component extends up to energies of at least 20~TeV~\cite{Vela}.
These results are already analyzed by the H.E.S.S. Collaboration,
to show that the state-of-the-art models for the high-energy emission of pulsars may be challenged~\cite{Vela}.
However, besides using these results to limit the models for pulsars, it is noteworthy that the photons from the Vela pulsar are very energetic, and the phasograms of the Vela pulsar have time structures of milliseconds.
Although this object is quite close to the earth, at a distance of 287~pc~\cite{Dodson:2003ai,Vela}, we suggest that the Vela pulsar can be used to constrain parameters of LIV, with the same method as that utilized in the case of GRBs, because of the photons of energies of at least 20~TeV and the high temporal variability of this source, which has a spin period of 89~ms~\cite{Vela}.
In the following, we perform a quick and rough analysis without more detailed data and constrain the leading order and next-to-leading order LIV parameters for photons.

It is speculated from quantum gravity that the Lorentz invariance might be broken at the Planck scale ($E_\mathrm{Pl}\simeq 1.22\times10^{19}~\mathrm{GeV}$),
and that the light speed may have a variation with the energy of the photon. 
For energy $E\ll E_{\rm Pl}$, the modified dispersion
relation (MDR) of the photon can be expressed to the leading order as
\begin{equation}\label{eq:MDR}
  E^2=p^2 c^2 \left[1-s_n\left(\frac{pc}{E_{\mathrm{LV,} n}}\right)^n\right].
\end{equation}
Assuming that the traditional relation $v=\partial E / \partial p$ holds, we have the following speed relation
\begin{equation}\label{eq:speed}
  v(E)=c\left[1-s_n\frac{n+1}{2}\left(\frac{pc}{E_{\mathrm{LV,}n}}\right)^n\right],
\end{equation}
where $n=1$ or $n=2$ as usually assumed, $s_n=\pm1$ indicates whether high-energy photons travel faster~($s_n=-1$)
or slower~($s_n=+1$) than low-energy photons, and $E_{\rm{LV},n}$ represents the nth-order Lorentz violation scale.
Usually, we should consider the expansion of the Universe~\cite{Jacob2008,Zhu:2022blp}.
However, since the Vela pulsar is located nearby at the distance of $L=287~{\rm pc}$, we can ignore the expansion of the Universe
and assume that the traveling time of cosmic particles can be written as $T=L/v(E)$.
Thus the arrival time difference between a high-energy photon with energy $E$ and a low-energy light  is
\begin{equation}\label{eq:dt}
\Delta T=\frac{L}{v(E)}-\frac{L}{c}=s_n \frac{(n+1)L}{2c}\left(\frac{E}{E_{\mathrm{LV,}n}}\right)^n,
\end{equation}
and the nth-order Lorentz violation scale can be written as
\begin{equation}\label{eq:elv}
E_{\mathrm{LV,}n}=\left(s_n\frac{(1+n)L}{2c\Delta T }\right)^\frac{1}{n}E.
\end{equation}

One aspect we need to notice is where are these ultra-high-energy photons produced, since the strong gravitational field close to the pulsar may influence the traveling time difference.
From Ref.[8], these ultra-high-energy photons are most likely created by the inverse-Compton (IC) scattering of high-energy electrons with low-energy photons.
To produce such high-energy photons, the energies of the electrons must be higher.
These electrons were accelerated for a long time in the electromagnetic field of the pulsar before the IC scattering,
which means that these ultra-high-energy photons are produced far away from the pulsar.
In other words, we can dismiss the influence of the gravitational field.  
During the propagation of the photons, the interstellar medium (ISM) may slow down the photons. 
But for ultra-high-energy photons, we can dismiss this effect because if ultra-high-energy photons interact with ISM, ultra-high-energy photons will be absorbed and create new particles.
From Fig.~1 of Ref.~\cite{Vela}, the phasograms of Vela as measured with H.E.S.S. CT1–4 for energies 
above 5~TeV and with H.E.S.S. CT5 in the 10 to 80~GeV range,
we can see a possible time difference of peak P2 between energy above 5~TeV and $10-80~\rm GeV$. 
We can see that the high-energy photons travel slower than low-energy photons, and that suggests $s_n=+1$.
The time difference between the two peaks is one-half of the bin width of the phasogram for energies above 5~TeV ($0.01\times$ spin period).
Considering that the time difference read from the phasograms can be also from bin division, 
which means that the actual time difference can be smaller, we can obtain that 
$\Delta T<0.01 T_{\rm period}=0.89~\rm ms $.
Since $5~{\rm TeV} \gg 80 ~{\rm GeV}$, we can ignore the contribution of the time difference from low-energy photons,
and the energy of high-energy photons has a relation $E>5~\rm TeV$. 
From Eq.~(\ref{eq:elv}), for $n=1$, we have
\begin{equation}
E_{\mathrm{LV,}1}=\frac{L}{c \Delta T}E > 1.66\times 10^{17} ~\rm GeV.\label{elv1}
\end{equation}
For $n=2$, we have
\begin{equation}
E_{\mathrm{LV,}2}=\left(\frac{3L}{2c\Delta T }\right)^\frac{1}{2}E>3.53\times 10^{10} ~\rm GeV.\label{elv2}
\end{equation}

It is worth noting that, the results we obtained in Eq.~(\ref{elv1}) and Eq.~(\ref{elv2}) are somewhat weaker compared to some existing constraints in the literature~\cite{wei2021tests,He:2022gyk} (see, also, a conservative bound $E_{\mathrm{LV},1} \ge 3.6 \times 10^{17} ~\mathrm{GeV}$ in Refs.~\cite{Shao2010f,Zhang2015,Xu2016a,Xu2016,Amelino-Camelia2017,Xu2018,Liu2018,Li2020,Zhu2021a,Chen2021}).
However these results can be expected to be refined by analyzing the data of the Vela pulsar more thoroughly and utilizing potentially more observations of very-high-energy photons from pulsars in the future.
We anticipate that more stringent constraints on Lorentz invariance violation parameters can be obtained.
On the other hand, the constraints obtained in this work, albeit weak, reveal the feasibility of utilizing pulsar photons to constrain Lorentz invariance violation parameters.
Therefore in this work, we show that the Vela pulsar expands the range of sources that can be chosen to effectively study Lorentz invariance violation.

Our approach can be also extended to analyze the traveling time difference for photons with very big energy differences from the Crab pulsar,
which is the most powerful pulsar in our Galaxy.  
The distance of the Crab pulsar is 6.5 times that of the Vela pulsar, 
while the Crab pulsar spins 30 times a second, with a spin period around 33~ms, which is comparable with 89~ms of the Vela pulsar.
Therefore we can estimate that the Crab pulsar may provide stronger or comparable constraints on the Lorentz violation parameter of photons
if there are data of around 10~TeV photons from Crab pulsar. Unfortunately at the moment, we have no information about 10-TeV scale photons from the Crab pulsar for our analysis, what we know is that the pulsed emission up to 1.5 TeV has been detected from the Crab pulsar~\cite{MAGIC:2015ggt}. With the constraints of Eqs.~(\ref{elv1}) and (\ref{elv2}), we can provide a rough formula that the predicted time delay between high-energy photons and low-energy photons satisfies $\Delta T<5.8~ \mathrm{ms}\times\left(
{E}/{5 ~\mathrm{TeV}}\right)^n$, where $E$ is the energy of high-energy photons and $n=1, 2$ is the order of Lorentz violation.

\section*{Acknowledgements}
This work is supported  
by National Natural Science Foundation of China under Grants No.~12335006 and No.~12075003.

\bibliographystyle{elsarticle-num} 
\bibliography{ref}






\end{document}